\documentclass[useAMS,usenatbib]{mn2e}
\usepackage{graphicx}
\usepackage{rotating}
\usepackage{longtable}
\usepackage{lscape}
\usepackage{amssymb,amsmath}
\def\lsim{\mathrel{\rlap{\lower 3pt \hbox{$\sim$}} \raise 2.0pt \hbox{$<$}}}
\def\gsim{\mathrel{\rlap{\lower 3pt \hbox{$\sim$}} \raise 2.0pt \hbox{$>$}}}

\def\msun{{\rm M}_\odot}
\def\mhost{{\rm M}_{\rm host}}
\def\mgal{{\rm M}_{\rm gal}}
\def\mbh{{\rm M}_{{\rm{BH}}}}
\def\civ{{\rm C\,{\sc IV}}}
\def\mgii{{\rm Mg\,{\sc II}}}

\def\feii{{\rm Fe\,{\sc II}}}
\def\oii{{\rm O\,{\sc II}}}
\def\hi{{\rm H\,{\sc I}}}
\def\qsof{QSO$_{\rm F}$}
\def\qsob{QSO$_{\rm B}$}
\def\sn{{\rm S/N}}
\def\pd{{\rm pd}}
\def\z{{\rm z}}
\def\zf{{\rm z}_{\rm F}}
\def\zb{{\rm z}_{\rm B}}
\def\ewr{EW$_{\rm r}$}

\title[On the cool gaseous haloes of quasars]
{On the cool gaseous haloes of quasars}
\author[Farina et al.]{
	E.~P.~Farina$^{1,2}$\thanks{E--mail: {\tt emanuele.farina@mib.infn.it}}, 
        R.~Falomo$^{3}$,
	R.~Decarli$^{4}$,
	A.~Treves$^{1,2,5}$, and
	J.~K.~Kotilainen$^{6}$\\
       	$^{1}$ Universit\`{a} degli Studi dell'Insubria, via Valleggio 11, I-22100 Como, Italy\\
	$^{2}$ INFN Milano--Bicocca -- Universit\`{a} degli Studi di Milano--Bicocca, Piazza della Scienza 3, I-20126 Milano, Italy\\
       	$^{3}$ INAF -- Osservatorio Astronomico di Padova, Vicolo dell'Osservatorio 5, I-35122 Padova (PD), Italy\\
	$^{4}$ Max-Planck-Institut f\"ur Astronomie, K\"onigstuhl 17, D-69117 Heidelberg, Germany\\
	$^{5}$ Associated to INAF\\
       	$^{6}$ Finnish Centre for Astronomy with ESO (FINCA), University of Turku, V\"ais\"al\"antie 20, FI-21500 Piikki\"o, Finland
	}
\begin{document}

\date{ }

\pagerange{\pageref{firstpage}--\pageref{lastpage}} \pubyear{2012}

\maketitle

\label{firstpage}

\begin{abstract}
	We present optical spectroscopy of projected QSO pairs to investigate the 
	\mgii\ and the \civ\ absorption features imprinted on the spectrum
	of the background object by the gaseous halo surrounding the foreground QSO. 
	We observed 13 projected pairs in the redshift range $0.7\lsim\zf\lsim2.2$ 
	spanning projected separations between 60\,kpc and 120\,kpc. 
	In the spectra of the background QSOs, we identify \mgii\ intervening 
	absorption systems associated to the foreground QSOs in 7 out of 10 pairs, 
	and~1~absorption system out of 3 is found for \civ.
	The distribution of the equivalent width as a function of the impact parameter 
	shows that, unlike the case of normal galaxies, some strong absorption 
	systems (\ewr$>$1\,\AA) are present also beyond a projected radius of $\sim70$\,kpc. 
	If we take into account the mass of the galaxies as an additional parameter that 
	influence the extent of the gaseous haloes, the distribution of the absorptions 
	connected to the QSOs is consistent to that of galaxies. In the spectra 
	of the foreground QSOs we do not detect any \mgii\ absorption lines originated by the 
	gas surrounding the QSO itself, but in 2 cases these features are present for \civ. 
	The comparison between the absorption features observed in the transverse 
	direction and those along the line of sight allows us to comment on the 
	distribution of the absorbing gas and on the emission properties of the 
	QSOs.\\ 
	
	Based on observations undertaken at the European Southern Observatory 
	(ESO) Very Large Telescope (VLT) under Programmes 085.B-0210(A) and
	086.B-0028(A).		 
\end{abstract}


\section{Introduction}\label{sec:1}

Absorption lines in quasar (QSO) spectra provide a unique tool for probing
the gas and dust content of foreground galaxies and of intergalactic medium 
at almost any redshift. 
Since the works of \citet{Bahcall1969} and \citet{Boksenberg1978} these 
features are considered to be linked to large gaseous haloes around galaxies 
extending up to $100$\,kpc \citep[see the review by][]{Churchill2005}.
Despite a number of studies aimed at connecting metal absorption systems 
(in particular \mgii\ and \civ) with galaxy properties 
\citep[e.g.,][]{Young1982, Lanzetta1987, Bergeron1991, Steidel1992, 
Steidel1994, Churchill1999, Rigby2002, Nestor2005, Bergeron2011}, the origin 
of the absorbing gas is still unclear. 
However, in the last decades, it is progressively emerging a scenario in which 
the stronger absorption lines (rest frame equivalent width \ewr$\gsim1$\,\AA) 
are linked to winds driven by the star formation activity, while the weaker one are 
associated with inflows of gas onto the host galaxies.

For instance, \citet{Zibetti2007}, stacking SDSS images of 2800 \mgii\ 
absorbers with $0.37<\z<1.00$, show that the stronger absorption systems 
reside preferentially in blue star--forming galaxies, while the weaker 
ones in red passive galaxies.
This result was confirmed by \citet{Menard2011} in a sample of 8500 \mgii\ 
absorption systems present in the SDSS QSO spectra. They discovered a strong 
correlation between the \ewr\ of the absorbers and the luminosity of the 
associated [\oii] lines, considered as a tracer of the star formation rate 
(SFR) of galaxies. \citet{Prochter2006} investigated the evolution of the 
number density of $\sim7400$ \mgii\ strong absorption systems and found
rough correspondence with the evolution of the star formation rate at redshifts 
between $\z=0.5$ and $2$, suggesting a link between the two phenomena.
Recently \citet{Nestor2011} reported clear evidence that ultrastrong \mgii\ 
absorbers (\ewr$\gsim3$\,\AA) reside in galaxies with very high SFRs.
In addition to that, spectroscopic observations of highly star forming galaxies 
reveals the presence of blueshifted \mgii\ absorption features, further supporting
the outflowing winds scenario \citep{Tremonti2007, Weiner2009, Rubin2010, 
Martin2012}.

The existence of a link between galaxy luminosity/halo mass and strength of 
absorption features, expected in the infalling scenario, is still debated. 
\citet{Chen2001} and \citet{Chen2010a} found that the \ewr\ and the extent of 
\mgii\ and \civ\ absorption systems are related to host galaxy luminosity.
More directly, \citet{Chen2010a} searched for absorption systems in a sample
of 94 galaxies at redshifts between $\z=0.1$ and $0.5$ within
$\pd\lsim85$\,kpc from a QSO sightline. The \ewr\ of the systems found 
(the great majority of which have \ewr$\lsim1$\,\AA) scales with the stellar 
mass and only little with the  star formation rate of the host galaxies
\citep{Chen2010b}. 
These results are interpreted considering that \mgii\ absorption systems 
arise from infalling clouds to fuel star formation.
\citet{Kacprzak2011} directly compared the relative \mgii\ halo gas and host
galaxy kinematics for 13 L$^{\star}$ galaxies at $\z\sim0.1$. They found 
that these galaxies have low SFRs and a kinematically quiescent
interstellar medium containing no outflowing gas. Given that these galaxies
live in isolated environment, they suggest a scenario in which the cool gas
halo was infalling and providing a gas reservoir that could maintain the
low levels of star formation within the host galaxies. 
\citet{Bowen2011}, in a sample of Luminous Red Galaxies (LRGs) with 
redshifts $0.46<\z<0.6$, found a low covering factor ($k$) for absorber 
with \ewr$\gsim0.6$\,\AA\ and a lack of correlation between \ewr\ and impact 
parameter or r--band absolute magnitude of the galaxies. 
They suggested that this is due or to the rich environment of these objects 
that makes their haloes too hot to maintain cool gas, or to the low 
rates of star formation that are not intense enough to fill their haloes 
with \mgii\ clouds.
Cross correlating \mgii\ absorption systems and the properties of LRGs 
\citet{Bouche2006} found at redshift $\z\sim0.5$ an anti--correlation between 
the halo mass and the \ewr\ of absorbers (most of which have \ewr$\gsim1$\,\AA), 
which suggests that the \mgii\ absorption systems are not virialised within 
the haloes. These results were further confirmed in larger sample by 
\citet{Lundgren2009} and by \citet{Gauthier2009}. Intriguingly, 
\citet{Rubin2012} detect inflow of gas into isolated star--forming galaxies 
at $\z\sim0.5$. 

In contrast to the large attention given to absorptions detected around
normal galaxies, only few studies have been focused on the properties of 
the gaseous halo of galaxies hosting a QSO in their centre.
The observation of projected QSO pairs allows to probe the properties 
of the foreground QSO (\qsof, $z \equiv \zf$), through the study of 
absorption features imprinted on the background QSO spectra (\qsob, 
$z \equiv \zb > \zf$). The standard model on the origin of 
QSO high luminosity requires an intense gas accretion on a supermassive black 
hole that dramatically increases its activity. Feasible mechanisms responsible 
for the gas infall are instabilities caused by strong gravitational 
interactions and galaxy mergers \citep[e.g.,][]{Canalizo2007, Bennert2008, 
Green2010}. 
The close (few hundreds kiloparsecs) environment of QSOs is expected to be 
populated by tidal debris, streams, and diffuse cool gas clouds, as commonly 
observed in interacting galaxies \citep[e.g.,][]{Sulentic2001, Cortese2006}.
Given their low surface brightness, most of the properties of these features 
could be investigated almost exclusively in absorption, especially at high 
redshift. 

Some studies have been performed to analyse the distribution of neutral 
hydrogen around QSOs. For instance \citet{Hennawi2006}, starting from a 
sample of 149 projected QSO pairs (projected distance: 
$30$\,kpc$\lsim {\rm pd} \lsim 2.5$\,Mpc; redshift: $1.8<\zf<4.0$), found that 
the probability to have an absorber with $N_{\rm HI}>10^{19}$\,cm$^{-2}$ 
coincident within 200\,kpc with a \qsof\ is high ($\sim50\%$), and that the 
distribution of these absorbers is highly anisotropic (\citealt{Hennawi2007},
see also \citealt{Kirkman2008, Prochaska2009}).
The study of projected QSO pairs gives also the possibility to investigate the 
so--called {\it transverse proximity effect}: i.e., the expected decrease of 
absorption systems in the Ly$\alpha$ forest of a \qsob\ due to the ionising 
flux of a \qsof. Intriguingly the many attempts to measure this effect,
with perhaps one exception \citep{Gallerani2008}, have led to only marginal
or no detection \citep[e.g.,][]{Crotts1989, Dobrzycki1991, Liske2001, 
Schirber2004}. The presence of a transverse proximity effect for heavier 
elements is still unclear. For instance it was observed for He\,{\sc II} by 
\citet{Worseck2006}, but not for \mgii\ \citep{Bowen2006} or for \civ\ 
\citep{Tytler2009}.

Up to now, only a few metal absorption systems associated or near to a 
QSO have been serendipitously discovered \citep[e.g.,][]{Shaver1982, 
Shaver1983, Shaver1985, Decarli2009}. The limited statistics is due to 
the little number of known close projected QSO pairs.
Even the large spectroscopic QSO catalogue of the Sloan Digital Sky
Survey, made of more than 100000 objects~\citep[][]{Schneider2010}, 
contains only 22 QSO pairs with separation less than~$15^{\prime\prime}$.
One should recall that the fiber collisions prevents the spectroscopic 
observations of two sources with separations below~$55^{\prime\prime}$
in a single plate.

\citet{Tytler2009} studied the distribution of metal absorption features 
(in most case \civ) present in the spectra of 170 close 
($\z_{\rm B}-\zf \lsim 0.5$) projected QSO pairs with 
separations $100$\,kpc$\lsim {\rm pd}\lsim 2.5$\,Mpc in order to investigate 
their Mpc scale clustering properties around QSOs and other absorbers.
They found 16 absorbers within $\pm500$~km/s from the \qsof, all with 
${\rm pd}>400$\,kpc, that cluster with an approximately isotropic distribution, 
with a hint of an excess for systems with redshift $\z<\zf$.

\citet{Bowen2006} select from the sample of projected QSO systems 
of \citet{Hennawi2006} 4 the pairs with $30\lsim{\rm pd}\sim100$\,kpc and 
$0.5\lsim \zf \lsim 1.5$. In all these systems they detect \mgii\ absorption 
lines clearly associated to \qsof\ (see Figure~\ref{fig:1}).
Despite the small sample size, they speculate that the absence of corresponding 
absorption lines in the spectra of \qsof\ involve a non isotropic distribution of 
the absorbing gas. On the basis of these data \citet{Chelouche2008} proposed that 
the gas in the outer region of QSOs (i.e., at radius larger than $\sim20$\,kpc) 
is distributed in the same way of ${\rm L}^\star$ galaxies, but their thermal 
and ionisation structure is highly influenced by the central black hole 
emission.

In this paper we search for \mgii\ and \civ\ absorption features in a 
sample of 13 projected pairs.
The aim of this work is to investigate the physical properties and the 
spatial distribution of the cool gas associated to the halo of QSOs. 
It is worth noting that the considered ions have very different ionisation 
energies, probing different gas conditions. 
The properties of the detected features are compared with those found in 
{\it normal} galaxies (i.e., galaxies not hosting a QSO) in order to 
investigate the effects of the central black hole emission on the cool gas.

We present our sample in \S\ref{sec:2} and the data reduction and analysis 
in \S\ref{sec:3}. In \S\ref{sec:4} we investigate the 
properties of the detected absorption systems. In \S\ref{sec:5} we 
compare and contrast our results with those found in galaxies. 
We conclude and summarise in \S\ref{sec:6}.

Throughout this paper we consider a concordance cosmology with 
H$_0=70$\,km/s/Mpc, $\Omega_{\rm m} = 0.3$, and $\Omega_\Lambda=0.7$.

\section[]{The sample}\label{sec:2}

We searched in the \citet{Veron2010} quasar catalogue for projected pairs that 
match the following requirements: 
(i) the angular separation is $\Delta \theta \lsim 15^{\prime\prime}$ that,
at the redshifts considered in our sample (i.e., $0.7\lesssim \zf \lesssim 
2.2$), allows us to explore the outer regions of a typical galaxy halo
(i.e., from 60~to~120\,kpc);
(ii) the redshifts of fore-- and back--ground QSOs combine so that the \civ\ and
\mgii\ emissions of the \qsof\ fall within the wavelength range observed 
with FORS2 GRISM~1400V or GRISM~1200R, avoiding important sky features or 
narrow emission lines in the spectrum of the \qsob; 
(iii) targets have declination $<15^{\circ}$, so that they are visible from the 
Paranal site; 
and finally (iv) the QSO$_{\rm B}$ is brighter than m$_{\rm V}\sim21$ in order
to collect spectra with signal--to--noise $\sn\gsim10$.
The resulting sample consists of 19 close pairs, 13 of which were observed with 
FORS2 at VLT.

\begin{figure}
\centering
\includegraphics[width=1.0\columnwidth]{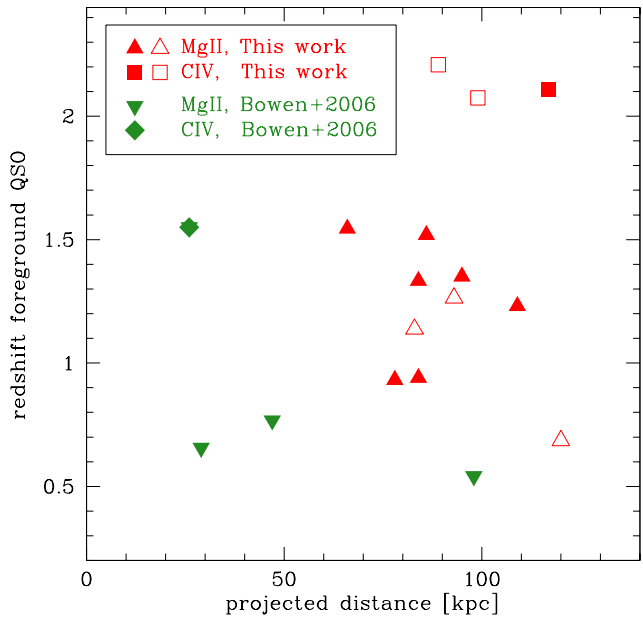}
\caption{Distribution of close projected pairs in the pd--$\zf$ plane.
Triangles are objects in which we investigate for the \mgii\ absorption features,
while squares for the \civ. The systems showing an absorption feature in the 
\qsob\ spectra associated to the \qsof\ are marked with filled points, otherwise 
with empty ones. Red and green points are from our sample and from that of 
\citet{Bowen2006}, respectively. The systems discovered by 
\citep{Tytler2009} have projected separations larger than $400$\,kpc and thus are 
not included in this Figure.}\label{fig:1}
\end{figure}

In Table~\ref{tab:1} and in Figure~\ref{fig:1} we present the general
properties of the 13~observed pairs. These are radio quiet QSOs, with an 
average angular separation of 11.4$^{\prime\prime}$ that corresponds to an average 
projected distance of $\sim90$\,kpc. In our sample we do not take into account 
the presence of absorption {\it a priori}, thus it seems suitable to estimate 
the unbiased frequency of absorption systems.

\begin{table*}
\centering
\scriptsize
\caption{Properties of the observed QSO projected pairs: 
most common name of the foreground QSO, 
our identification label of the pair (ID), 
redshift from broad emission line ($\z_{\rm bl}$), 
absolute V--band magnitude of the QSO (V),
angular ($\Delta \theta$) and projected (pd) separation,
 bolometric luminosity (L$_{\rm bol}$), black hole mass 
(M$_{\rm BH}$, see text for details), average seeing 
during observations (See.), and average signal--to--noise 
ratio per pixel on the continuum close to the expected
position of the absorption lines (S/N, see \S\ref{sec:4}).
The label ${\rm F}$ and ${\rm B}$ refer to the foreground
and to the background QSO, respectively.
For notes on the individual objects see 
Appendix~\ref{app:1}. 
}\label{tab:1}
\begin{tabular}{llccccccccccc}
\hline
\qsof              &  ID   & $\z_{\rm bl,F}$ & $\z_{\rm bl,B}$ & V$_{\rm F}$ & V$_{\rm B}$ & $\Delta\theta$ & pd	& L$_{\rm bol,F}$    & M$_{\rm BH,F}$		 & See.   & S/N$_{\rm F}$ & S/N$_{\rm B}$ \\
	           &       &		          &		         &	       &  	     & [arcsec]	      & [kpc]   & [$10^{46}$\,erg/s] & [$\times 10^8\,\msun$]	 & [arcsec] &		  &		  \\	 
\hline  		 
\hline  		 
SDSSJ00022-0053B   &  QQ01 & 1.542 & 2.205 & -24.83 & -26.66 & \phantom{1}7.8 & \phantom{1}66  & 0.89$\pm$0.04 &           12.3 & 0.71 & 12 & 13\\
2QZJ003954-2725C   &  QQ02 & 1.262 & 2.100 & -24.29 & -23.96 &           11.1 & \phantom{1}93  & 0.56$\pm$0.01 & \phantom{1}6.2 & 0.84 & 20 & 10\\
2QZJ004344-3000B   &  QQ03 & 1.346 & 1.554 & -22.67 & -24.71 &  	 11.3 & \phantom{1}95  & 0.27$\pm$0.03 & \phantom{1}4.6 & 0.88 & 10 & 14\\
SDSSJ00541-0946B   &  QQ04 & 2.113 & 2.113 & -25.59 & -28.05 &  	 14.1 & 	  117  & 1.89$\pm$0.02 &           13.7 & 1.41 & 49 & 20\\
Q0059-2702B        &  QQ05 & 0.941 & 1.963 & -24.96 & -23.95 &  	 10.7 & \phantom{1}84  & 2.23$\pm$0.06 &           10.4 & 1.35 & 41 & 22\\
2QZJ011050-2719    &  QQ06 & 1.332 & 2.254 & -23.59 & -24.86 &  	 10.0 & \phantom{1}84  & 0.60$\pm$0.02 & \phantom{1}6.8 & 1.36 & 23 & 10\\
2QZJ101636-0234A   &  QQ07 & 1.518 & 3.448 & -25.65 & -26.64 &  	 10.1 & \phantom{1}86  & 2.11$\pm$0.13 & \phantom{1}9.9 & 0.75 & 15 & 15\\
2QZJ102425+0013A   &  QQ08 & 1.138 & 2.350 & -23.99 & -26.13 &  	 10.1 & \phantom{1}83  & 0.21$\pm$0.05 & \phantom{1}0.9 & 1.18 &  6 & 9 \\
SDSSJ11318-0222A   &  QQ09 & 2.198 & 2.353 & -25.65 & -26.95 &  	 10.8 & \phantom{1}89  & 0.54$\pm$0.11 & \phantom{1}6.5 & 2.35 &  5 & 5 \\
SDSSJ22028+1236A   &  QQ10 & 2.063 & 2.504 & -26.80 & -26.35 &  	 11.8 & \phantom{1}99  & 4.09$\pm$0.11 &           24.4 & 0.89 & 31 & 16\\
SDSSJ22067-0039A   &  QQ11 & 1.230 & 1.516 & -25.32 & -25.60 &  	 13.1 & 	  109  & 1.41$\pm$0.09 &           10.3 & 0.71 & 16 & 11\\
2QZJ222446-3200    &  QQ12 & 0.689 & 0.731 & -23.32 & -22.99 &  	 17.0 & 	  120  & 0.34$\pm$0.03 & \phantom{1}1.3 & 0.79 & 29 & 18\\
Q2225-4023B        &  QQ13 & 0.931 & 2.398 & -23.51 & -26.24 & \phantom{1}9.9 & \phantom{1}78  & 0.26$\pm$0.02 & \phantom{1}3.5 & 0.99 & 20 & 22\\
\hline
\end{tabular}
\end{table*}

\section[]{Observations and data Analysis}\label{sec:3}

\subsection[]{Observations and data reduction}

Spectra of the projected QSO pairs were collected with the VLT Antu telescope at
the ESO Paranal observatory. Observations were performed with GRISM 1200R and 
1400V on the FOcal Reducer/low dispersion Spectrograph 
\citep[FORS2, see][]{Appenzeller1998}, yielding a spectral resolution of 
R(1200R)=2900 and R(1400V)=2800 (with the 1$^{\prime\prime}$ slit). Typical exposure times 
are $\sim4000$~seconds ($\sim6000$~seconds for fainter objects). The only exceptions 
are QQ08 and QQ09, for which exposures no longer than 2600~seconds were allowed
by poor weather conditions.

Standard \texttt{IRAF}\footnote{\texttt{IRAF} is distributed by the National 
Optical Astronomy Observatories, which are operated by the Association of 
Universities for Research in Astronomy, Inc., under cooperative agreement 
with the National Science Foundation.} tools were used in the data reduction. 
The \texttt{ccdred} package was employed to perform bias subtraction, flat 
field correction, image alignment and combination. Cosmic rays were eliminated 
by combining different exposures and by applying the \texttt{crreject} 
algorithm. The spectra extraction, the background subtraction, and the 
calibrations both in wavelength and in flux were performed with the 
\texttt{twodspec} and \texttt{onedspec} packages. Our accuracy in the 
wavelength calibration is $\sim0.1$\,\AA. Galactic extinction was 
accounted for according to \citet{Schlegel1998}, assuming R$_{\rm V}=3.1$. 
The spectra obtained are presented in Figure~\ref{fig:2}.

\begin{figure*}
\centering
\includegraphics[width=2.0\columnwidth]{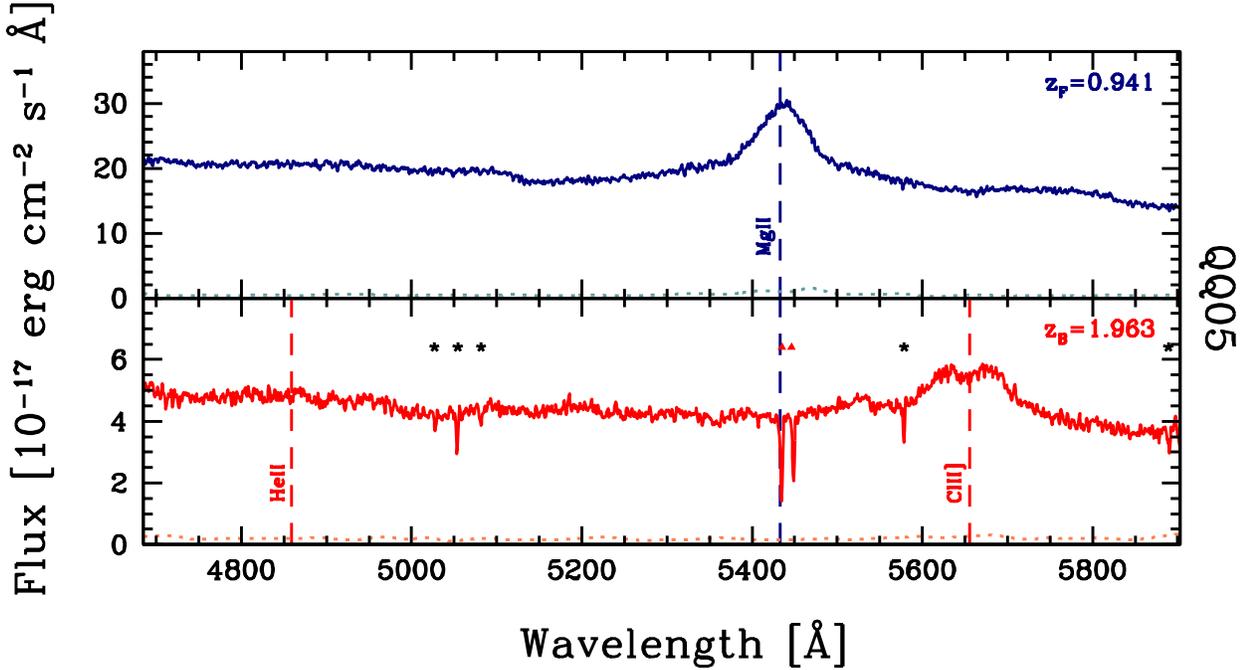}
\caption{Spectra of the projected QSO pairs corrected for Galactic extinction. 
The blue solid lines refer to \qsof\ and the red ones to \qsob. The 
corresponding $1\sigma$ error spectrum is shown at the bottom of each panel. 
Triangles and squares point to all the \mgii\ and \civ\ absorptions identified, 
respectively, while stars to other lines detected above a 3$\sigma$ threshold 
(see~\S\ref{sec:3.3}, Table~\ref{tab:2} and Table~\ref{tab:3}). Main QSO 
emission lines are labelled and the gray regions mask regions with prominent 
telluric features. Figures for the remaining pairs are available only
in the electronic edition of MNRAS.}\label{fig:2}
\end{figure*}

\addtocounter{figure}{-1}
\begin{figure*}
\centering
\includegraphics[width=2.1\columnwidth]{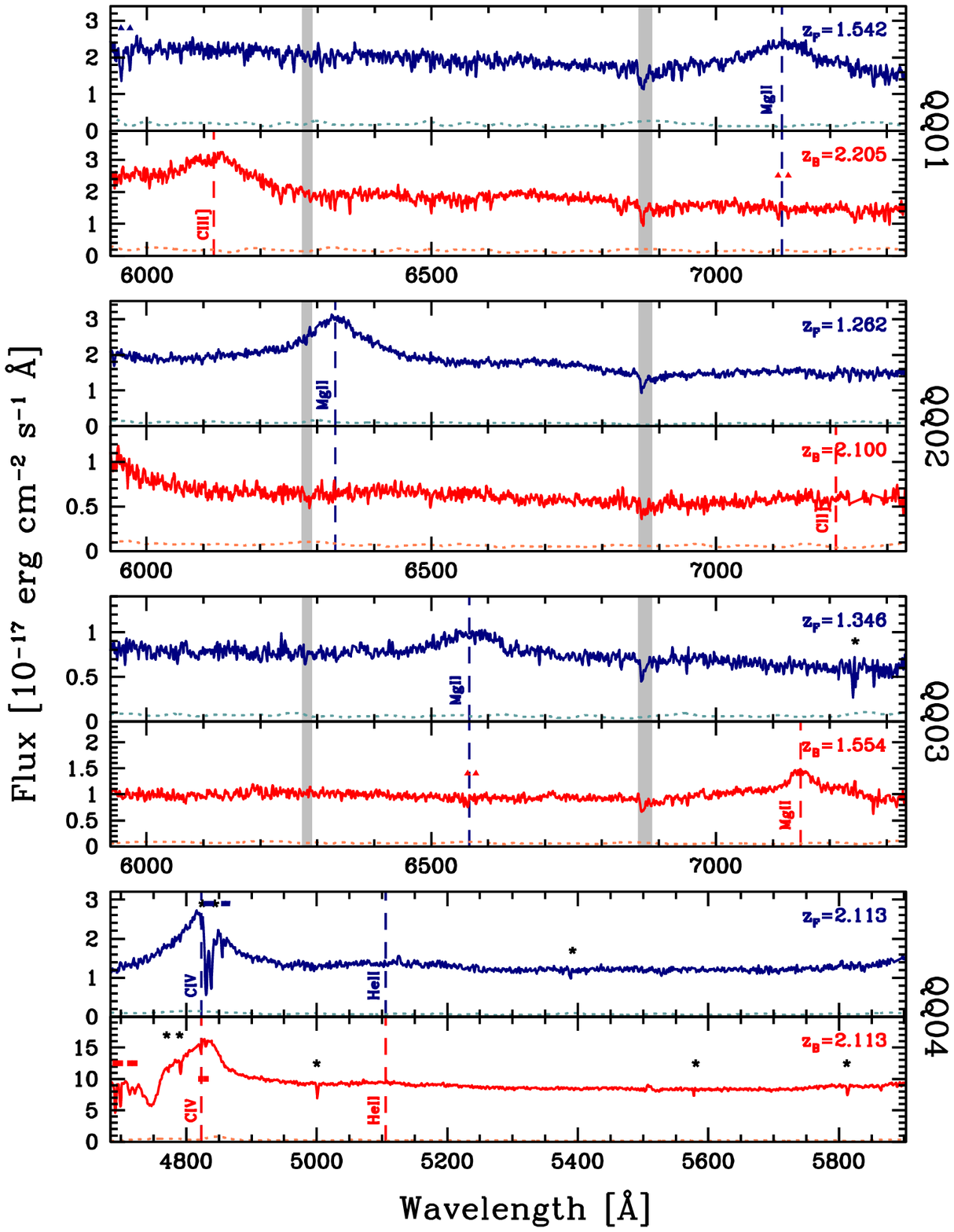}
\caption{ continued.}
\end{figure*}

\addtocounter{figure}{-1}
\begin{figure*}
\centering
\includegraphics[width=2.1\columnwidth]{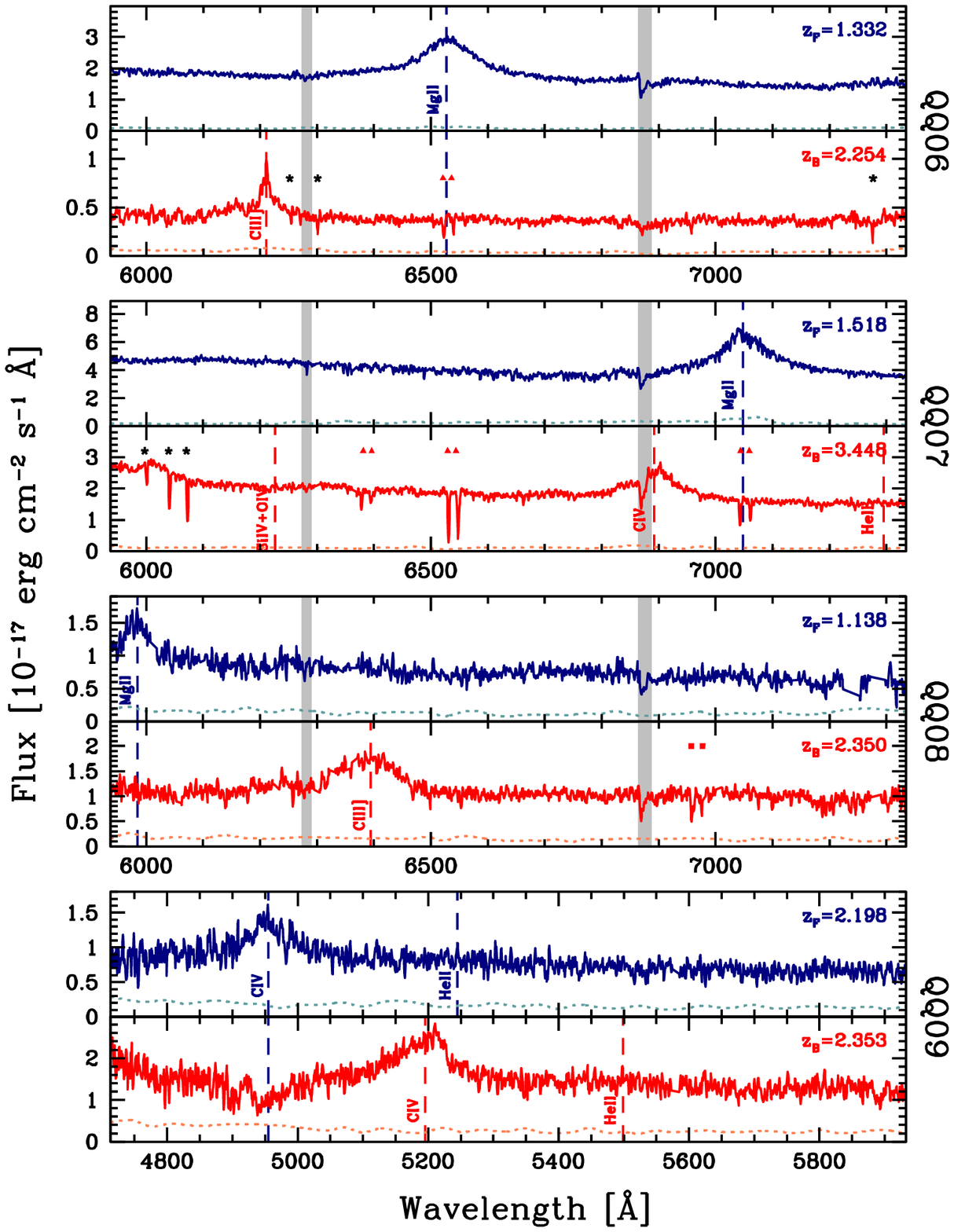}
\caption{ continued.}
\end{figure*}

\addtocounter{figure}{-1}
\begin{figure*}
\centering
\includegraphics[width=2.1\columnwidth]{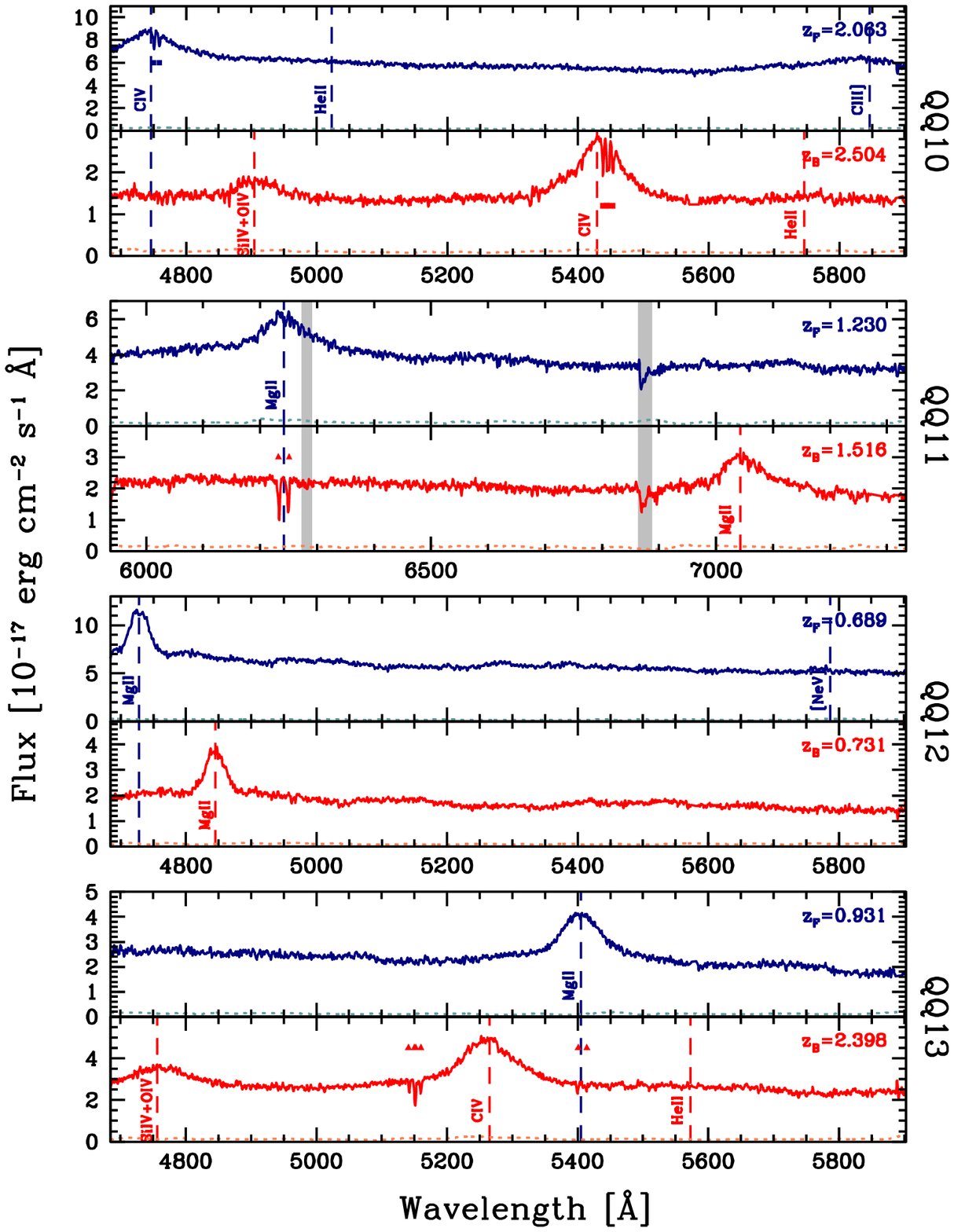}
\caption{ continued.}
\end{figure*}

\subsection[]{Analysis of the QSO Spectra}\label{sec:3.2}

For the analysis of the QSOs spectra we followed the procedure presented in 
\citet{Decarli2010a} and \citet{Derosa2011}. Namely, data are first inspected 
by eyes and regions showing apparent absorption features are masked. Then, we 
designed the continuum with the superposition of: 
(i) the non thermal power--law--like component; 
(ii) the host galaxy star light (assuming the elliptical galaxy template of 
\citealt{Mannucci2001});
and (iii) the contribution from blended \feii\ multiplets 
(modelled with template of \citealt{Vestergaard2001} in the UV band and 
with our original spectrum of the narrow line Seyfert 1 galaxy 
I~Zwicky~1 in the optical band).
Finally, we have fitted broad emission lines with two Gaussian curves with the 
same central wavelength \citep[see][]{Decarli2008}. Uncertainties on derived 
quantities are estimated from the $1\sigma$ errors in both continuum and line 
fits.

\subsection[]{Analysis of the Absorption Features}\label{sec:3.3}

Since the width of the absorption lines present in the spectra are smaller 
than the instrumental profile (${\rm FWHM}\sim100$\,km/s), for the analysis 
of the absorption features we considered a method similar to that adopted in 
the HST QSO Absorption Line Key Project \citep[][]{Schneider1993}.

We split the spectrum of QSOs (after masking the apparent absorption features) 
in intervals of fixed size \citep[20\,\AA, see for a similar 
approach][]{Sbarufatti2005}, and we modelled the ``continuum'' by interpolating 
the median values of the flux in each bin with a cubic spline function.
The equivalent width (EW) of an unresolved absorption feature and the relative 
uncertainties were calculated by modelling the Instrumental Spread Function 
(ISF) with a Gaussian with FWHM equal to the spectral resolution. From the 
redshift of the foreground QSOs, we are able to estimate the expected location 
of \mgii\ and \civ\ absorption features (see~\S\ref{sec:4}), thus a 3$\sigma$ 
threshold imposed on the detection \mgii($\lambda2796$) and \civ($\lambda1548$)
lines seems adequate to identify the absorption systems 
\citep[e.g.,][]{Churchill2000}.

The EWs and the centroid positions were measured by fitting a single Gaussian 
function \citep[e.g.,][]{Churchill2000} on the detected  absorption lines with 
our own software that performs a $\chi^2$ minimisation procedure. The $1\sigma$ 
uncertainties were estimated through standard error propagation and assuming that 
largest source of errors is given by the continuum placement.

Figures~\ref{fig:3} and \ref{fig:4} show the results of the fitting procedure 
on the detected absorption systems. Notes on the analysis of individual 
objects are in Appendix~\ref{app:1}.

\section[]{Absorption systems associated to QSOs}\label{sec:4}

We estimate the covering factor ($k$) of cool gas around QSOs basing on the 
detection of absorbers close to the redshift of the \qsof. 
It is well known that the redshifts derived from broad emission lines can 
differ from the systemic redshift by even hundreds km/s 
\citep[e.g.,][]{Tytler1992, Bonning2007}, and that absorbers within up to thousands km/s from a QSO are 
still connected with the QSO itself \citep[e.g.,][]{Wild2008}. 
Thus, as an operational definition, we consider an absorption system as associated 
to a QSO if its velocity difference with the the broad line redshift is smaller than 
$\pm1000$\,km/s. 
Throughout the Paper, we will refer to {\it transverse} or to 
{\it line--of--sight} ({\it LOS}) absorption features associated to \qsof\ 
depending on whether they are observed in the the spectrum of \qsob\ or of 
\qsof, respectively.

We can not discriminate between absorbers associated to the \qsof\ and 
absorbers that occur by chance coincidence. However, the small separations 
in both projected distances and relative velocities suggest that we are 
probing the cool gas strictly related to the QSO. 

The properties of the detected absorption systems associated to 
the \qsof\ are listed in Table~\ref{tab:2}.

\begin{table*}
\centering
\caption{Properties of absorption features associated to the foreground QSO: 
our identification label of the QSO (ID), 
feature detected (Abs.),
observed wavelength ($\lambda_{\rm abs}$),
rest frame equivalent width (EW$_{\rm r}$),
doublet ratio (DR), and
redshift (z$_{\rm abs}$).
If no absorption system is present, the 2$\sigma$ upper limit
for the EW$_{\rm r}$ is quoted. 
The label ${\rm F}$ and ${\rm B}$ refer to absorption systems observed on the
spectra of the foreground and of the background QSO, respectively.
}\label{tab:2}
\begin{tabular}{lcccccc}
\bf{Mg\,II} &&&&& \\      	  
\hline
ID     &  $\lambda_{\rm abs}(\lambda2796)$ & EW$_{\rm r}(\lambda2796)$ &  $\lambda_{\rm abs}(\lambda2803)$ & EW$_{\rm r}(\lambda2803)$ & DR & z$_{\rm abs}$ \\ 
       &  [\AA]	                           & [\AA]        	       &  [\AA] 			   & [\AA]                     &    &              \\      	  
\hline  		 
\hline  		 
 QQ01F & \dots   & $<0.28$       & \dots   & $<0.28$       & \dots          & \dots             \\
 QQ01B & 7109.0  & 0.48$\pm$0.17 & 7126.6  & 0.37$\pm$0.14 & 1.30$\pm$0.35  & 1.5421$\pm$0.0003 \\
\hline
 QQ02F & \dots   & $<0.20$       & \dots   & $<0.20$       & \dots          & \dots             \\
 QQ02B & \dots   & $<0.31$       & \dots   & $<0.31$       & \dots          & \dots             \\
\hline
 QQ03F & \dots   & $<0.28$       & \dots   & $<0.28$       & \dots          & \dots             \\
 QQ03B & 6564.1  & 0.46$\pm$0.12 & 6578.7  & 0.25$\pm$0.14 & 1.86$\pm$0.71  & 1.3470$\pm$0.0003 \\
\hline
 QQ05F & \dots   & $<0.14$       & \dots   & $<0.14$       & \dots          & \dots             \\
 QQ05B & 5434.5  & 1.21$\pm$0.09 & 5448.6  & 0.77$\pm$0.09 & 1.57$\pm$0.03  & 0.9434$\pm$0.0001 \\
\hline
 QQ06F & \dots   & $<0.19$       & \dots   & $<0.19$       & \dots          & \dots             \\
 QQ06B & 6522.8  & 0.91$\pm$0.17 & 6539.7  & 0.56$\pm$0.10 & 1.63$\pm$0.11  & 1.3326$\pm$0.0003 \\
\hline
 QQ07F & \dots   & $<0.23$       & \dots   & $<0.23$       & \dots          & \dots             \\
 QQ07B & 7043.0  & 0.58$\pm$0.07 & 7060.9  & 0.47$\pm$0.09 & 1.23$\pm$0.06  & 1.5186$\pm$0.0001 \\
\hline
 QQ08F & \dots   & $<0.41$       & \dots   & $<0.41$       & \dots          & \dots             \\
 QQ08B & \dots   & $<0.37$       & \dots   & $<0.37$       & \dots          & \dots             \\

\hline
 QQ11F & \dots   & $<0.22$       & \dots   & $<0.22$       & \dots          & \dots             \\
 QQ11B & 6233.4  & 1.36$\pm$0.16 & 6249.5  & 0.95$\pm$0.15 & 1.44$\pm$0.05  & 1.2291$\pm$0.0003 \\
\hline
 QQ12F & \dots   & $<0.15$       & \dots   & $<0.15$       & \dots          & \dots             \\
 QQ12B & \dots   & $<0.20$       & \dots   & $<0.20$       & \dots          & \dots             \\
\hline
 QQ13F & \dots   & $<0.19$       & \dots   & $<0.19$       & \dots          & \dots             \\
 QQ13B & 5399.6  & 0.24$\pm$0.07 & 5413.7  & 0.16$\pm$0.05 & 1.51$\pm$0.28  & 0.9310$\pm$0.0002 \\
\hline 
\hline
         &&&&& \\      	  
\bf{C\,IV} &&&&& \\      	  
\hline
ID     &  $\lambda_{\rm abs}(\lambda1548)$ & EW$_{\rm r}(\lambda1548)$ &  $\lambda_{\rm abs}(\lambda1551)$ & EW$_{\rm r}(\lambda1551)$ & DR & z$_{\rm abs}$ \\ 
       &  [\AA]	                           & [\AA]        	       &  [\AA] 			   & [\AA]                     &    &   	    \\         
\hline  		 
\hline 
 QQ04F & 4830.3  & 1.42$\pm$0.12 & 4838.1  & 1.36$\pm$0.12 & 1.04$\pm$0.02  & 2.1198$\pm$0.0003 \\
 QQ04B & 4822.4  & 0.12$\pm$0.02 & 4830.1  & 0.06$\pm$0.02 & 2.00$\pm$0.21  & 2.1147$\pm$0.0002 \\
\hline
 QQ09F & \dots   & $<0.39$       & \dots   & $<0.39$       & \dots          & \dots             \\
 QQ09B & \dots   & $<0.38$       & \dots   & $<0.38$       & \dots          & \dots             \\
\hline
 QQ10F & 4751.3  & 0.14$\pm$0.03 & 4759.1  & 0.13$\pm$0.07 & 1.08$\pm$0.35  & 2.0689$\pm$0.0005 \\
 QQ10B & \dots   & $<0.23$       & \dots   & $<0.23$       & \dots          & \dots             \\
\hline 
\hline
  										       
\end{tabular}
\end{table*}

\subsection{Transverse absorption systems}

\begin{figure*}
\centering
\includegraphics[width=1.99\columnwidth]{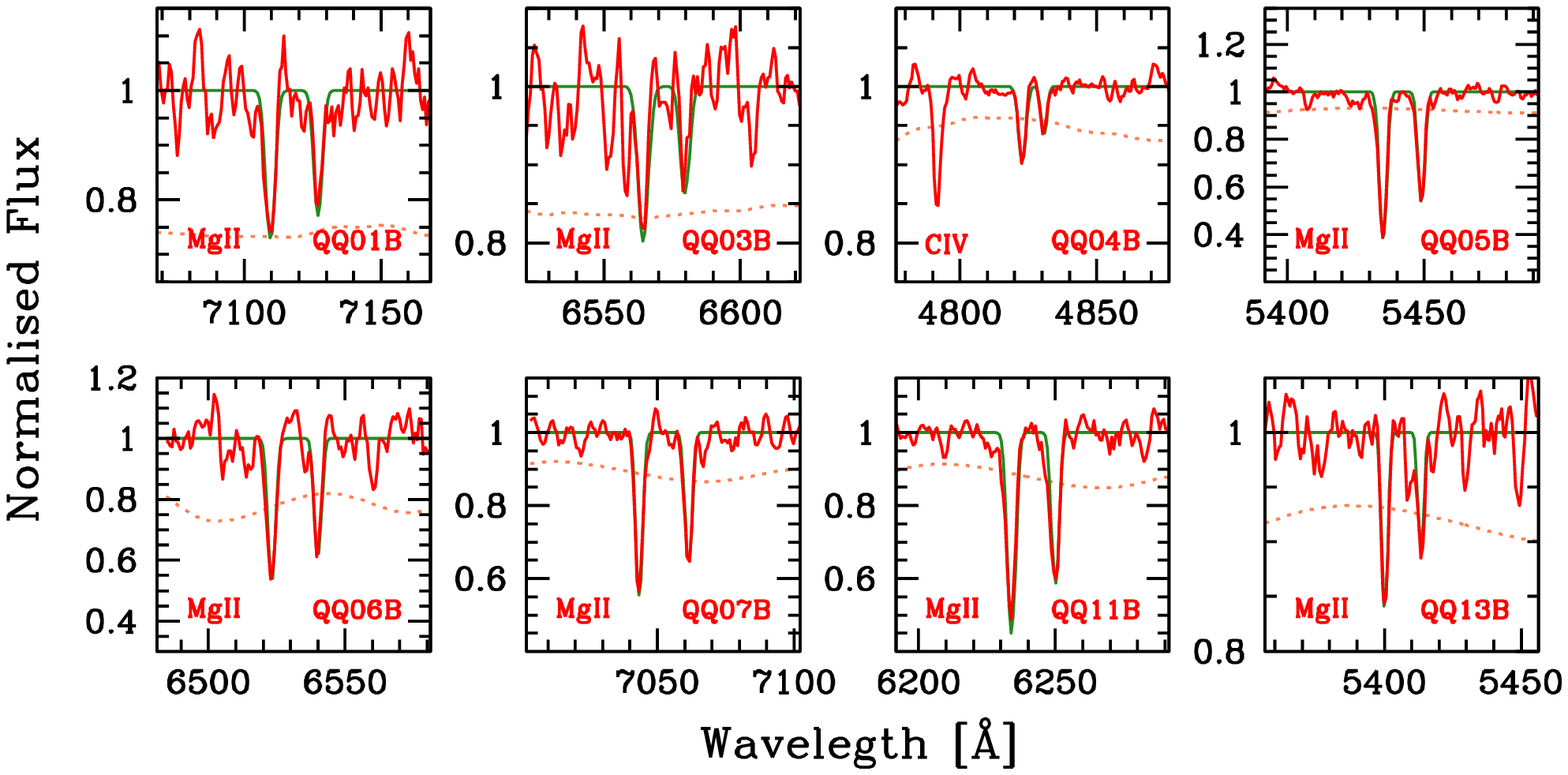}
\caption{Close up of the normalised \qsob\ spectra where the {\it transverse} 
absorption systems are detected (red line), the fit performed as
described in \S\ref{sec:3} (green line), and the 3$\sigma$ spectrum
(pale red dotted line). 
The identification of the feature at $\sim4791$\,\AA\ present in the spectra 
of QQ04B is uncertain due to the presence of the broad absorption line (see 
Appendix~\ref{app:1}). It seems not associated to any of the detected \civ\ 
absorption systems. 
}\label{fig:3}
\end{figure*}

\mgii\ transverse absorption features are present in 7 out of 10 pairs (see 
Figure~\ref{fig:3}). Through the doublet ratio method \citep[see e.g.,][]{
Chan1971, Falomo1990}, we can estimate an average column density of 
$\log \left(N_{\rm Mg\,II}/{\rm cm}^2\right)\sim 13.3$, assuming a Doppler 
parameter of $b\sim5$\,km/s \citep{Churchill1997}.

Excluding QQ08, for which our data do no allow to state significative 
upper limit to \ewr, we estimated covering fraction of 
$k({\rm Mg\,II})\sim67\%$ for 
systems with \ewr$>0.38$\,\AA. This value is lower than (but consistent 
within poissonian errors) the 100\% found by \citet{Bowen2006} in a sample of 
4 QSO pairs. Combining the two results we obtain $k({\rm Mg\,II})\sim77\%$ for 
\ewr$({\rm Mg\,II})>0.38$\,\AA. It is worth noting that, at higher redshift, 
\citet{Hennawi2006} found a similar high covering fraction for \hi\ asborbers 
($k({\rm H\,I})\sim75\%$) in QSO pairs with projected separations 
$\pd \lsim 150$\,kpc.
Moreover, if we consider the empirical relation between the \hi\ column 
densities and the \mgii\ equivalent widths estimated by \citet{Menard2009} in 
the sample of low--redshift Lyman absorbers of \citet{Rao2006}, we find that 
the covering fraction of $k({\rm H\,I})=50\%$ for absorbers with 
$\log(N_{\rm H\,I}/{\rm cm}^2)>19$ calculated by \citet{Hennawi2006}, exactly 
corresponds to the $k({\rm Mg\,II})=50\%$ for systems with \ewr$>0.5$\,\AA\ 
detected around low redshift QSOs.

The QSO \mgii\ covering factor seems to be larger than that observed in 
normal galaxies, although for these objects it spans a wide range of values, 
depending on impact parameter and galaxy properties: 
i.e., from $k$(\mgii)$\sim10\%$--$15\%$ \citep{Bowen2011}, 
to $25\%$ \citep{Bechtold1992}, 
to $\sim50\%$ \citep{Tripp2005,Kacprzak2008}, 
up to $\sim70\%$ \citep{Chen2010a}. 

Only QQ04B shows a strong \civ\ absorption feature close to $\zf$ (see 
Figure~\ref{fig:3}), formally yielding a covering 
factor of $k($\civ$)\sim30\%$. 
However, this feature is superimposed to the \civ\ broad emission
lines of the \qsob, and we can not exclude that it is instead a 
LOS absorption system associated to the \qsob\ (see Appendix~\ref{app:1}). 
In the literature only few cases of \civ\ transverse absorption systems have been
discovered, thus a sound value for the \civ\ covering factor around QSOs is 
still missing.
\citet{Bowen2006} have detected 
a \civ\ absorption system in correspondence of a \mgii\ one in a QSO pair separated by
a projected distance of 26\,kpc. In their sample \citet{Tytler2009} found 16 
\civ\ absorption features close in redshift to a QSO but both the small redshift difference 
and the large projected distance (pd$>$400\,kpc) of the QSO pairs they investigate 
do not allow to put constraints on the covering factor of \civ\ in the gaseous halo 
of the QSOs. Concerning galaxies, \citet{Chen2001} have shown that at impact 
parameter $\lsim 70$\,kpc the \civ\ covering factor is nearly unity.

\subsection{Line--of--Sight Absorption Systems}

\begin{figure}
\centering
\includegraphics[width=1.00\columnwidth]{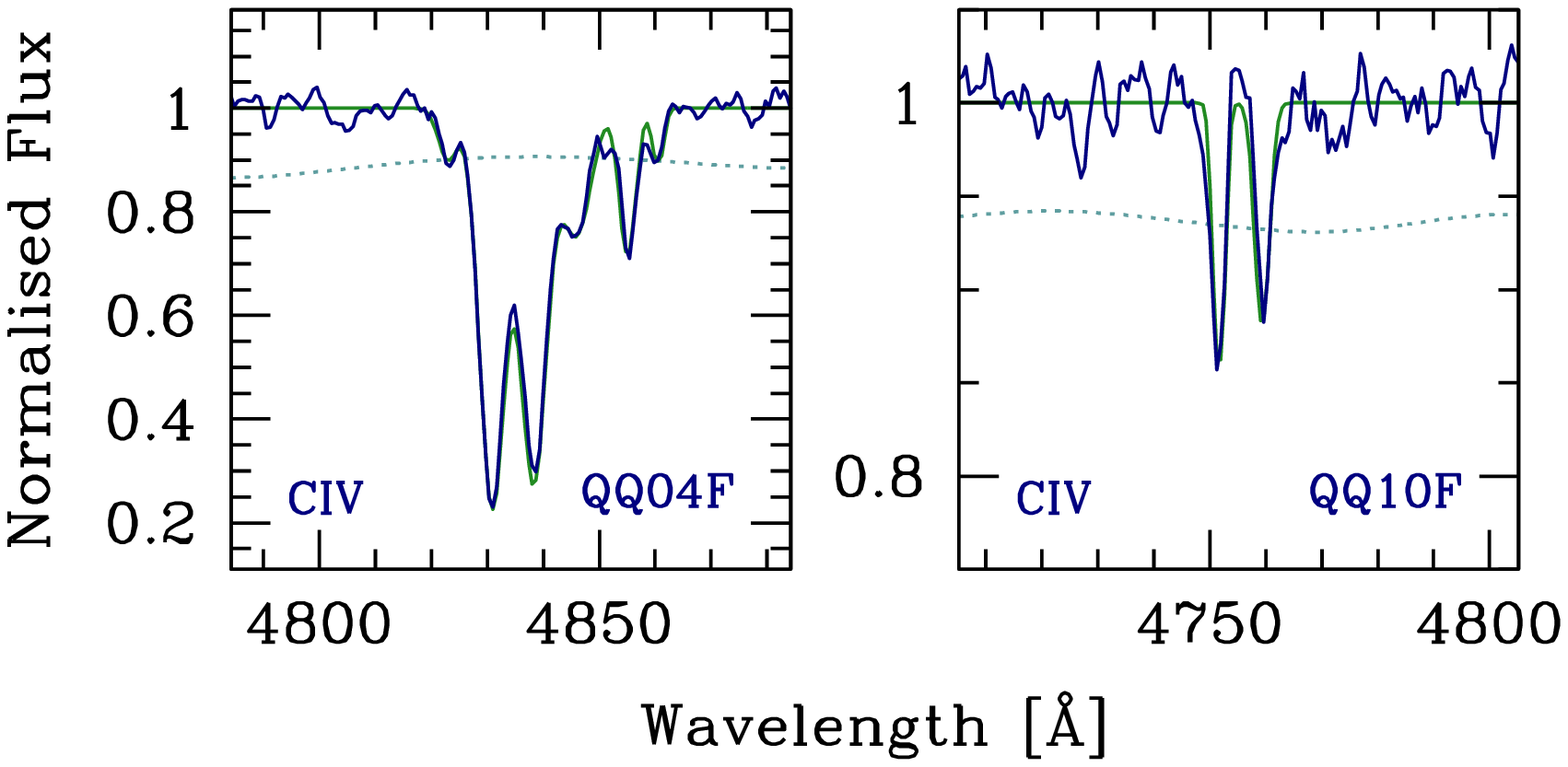}
\caption{Close up of the normalised spectra of the QSOs QQ04F and QQ10F for 
which the associated LOS absorption features are detected (blue line), the 
fit performed as described in \S\ref{sec:3} (green line), and the 
3$\sigma$ spectrum (pale blue dotted line). For notes on the 
different \civ\ doublets superimposed to the broad emission line of QQ04F see 
Appendix~\ref{app:1})}\label{fig:4}
\end{figure}

While no LOS absorption systems are observed for \mgii, these are present in 2 out 
of 3 of our spectra for \civ\ (in average 
$\log \left(N_{\rm C\,IV}/{\rm cm}^2\right)\sim13.7$, assuming a 
Doppler parameter of $14$\,km/s \citealt{Dodorico1998}, see Figure~\ref{fig:4}). 
This is in good agreement
with \citet{VandenBerk2008} and \citet{Shen2011a} who show that the occurrence
of \mgii\ associated LOS absorption features in SDSS spectra is only of a few percent,
and with \citet{Vestergaard2003} that investigate a sample of moderate redshift
QSO and found LOS narrow \civ\ absorption lines in $\gsim50\%$ of them
\citep[see also,][]{Wild2008}. We stress that in general we can
not discriminate between an origin of these absorbers close to the black hole,
or in its host galaxy or in its surrounding halo 
\citep[see for instance][]{Crenshaw2003}.

\section[]{Discussion}\label{sec:5}

\begin{figure}
\centering
\includegraphics[width=1.0\columnwidth]{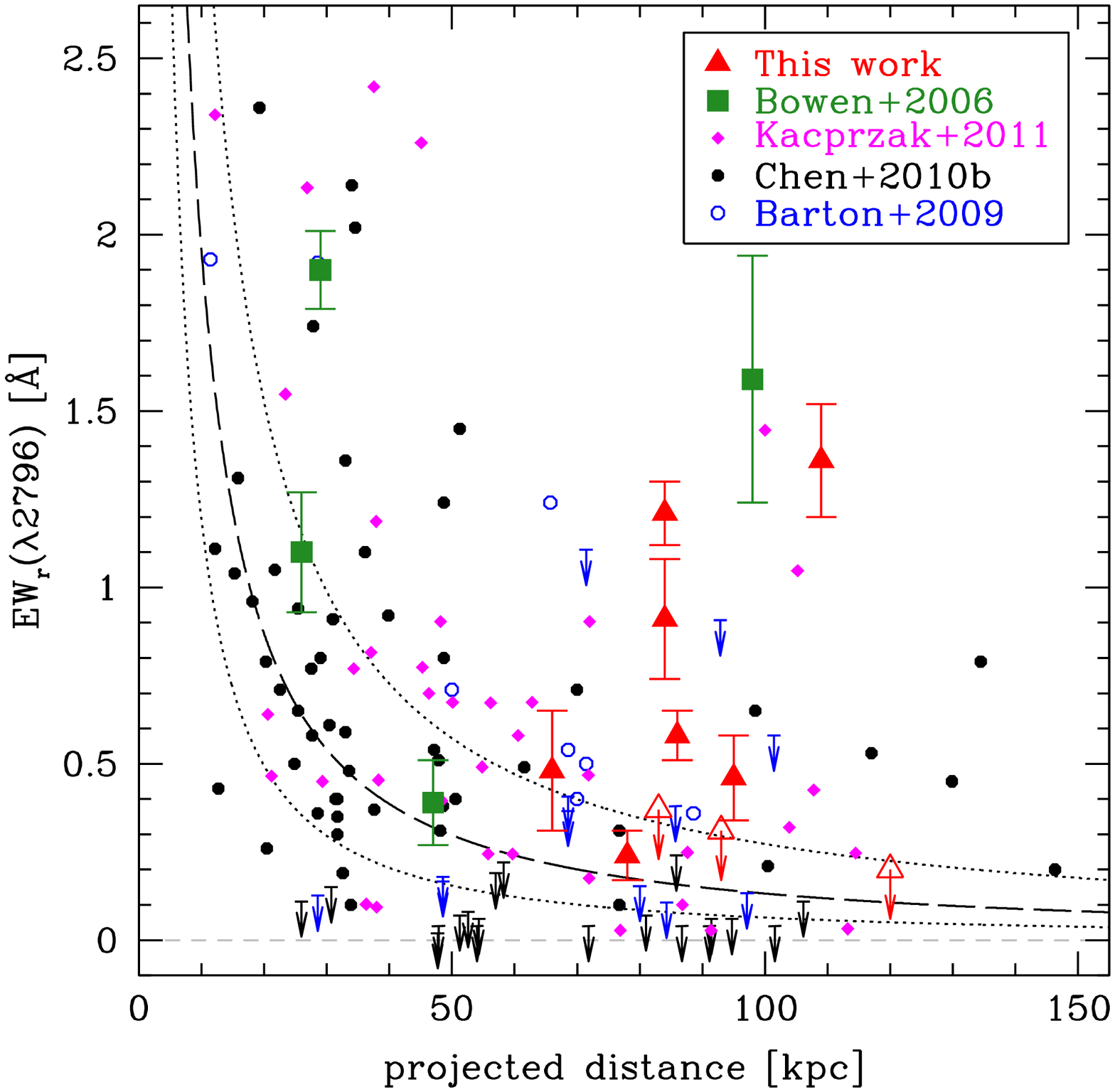}
\caption{Rest frame EW of \mgii($\lambda2796$) absorption line as a function of 
projected distance. Red filled triangle are the \qsof\ in which this 
feature is detected and the empty ones are the 2$\sigma$ upper limits. 
Green squares are data for QSOs from \citet{Bowen2006}. Magenta diamonds, 
black filled points, and blue empty circles are absorption features associated 
to galaxies from \citet{Kacprzak2011}, \citet[][also including the
absorption detected in {\it group galaxies}]{Chen2010a}, and 
\citet{Barton2009}. 
For the sake of comparison, the upper limits listed 
by~\citet[][]{Barton2009} were converted to the considered 2$\sigma$ limits.
Black dashed line shows the best fit of the 
anti--correlation proposed by \citet{Chen2010a} and the associated 
1$\sigma$ uncertainties 
(dotted lines).}\label{fig:5}
\end{figure}

In Figure~\ref{fig:5} we plot the rest frame equivalent width of
\mgii\ transverse absorption systems against the \qsof\ impact parameter. 
In spite of the small sample considered we note that many strong (i.e., 
\ewr$>$1\,\AA) absorption systems are located up to separations larger 
than 70\,kpc, in contrast to what observed in normal galaxies 
\citep[see e.g.,][]{Chen2010a}.
In this Section we consider the host galaxy mass as an additional
parameter that affects the \ewr--pd anti--correlation and we study
how the QSO emission influence the distribution of the cool 
gas. In \S\ref{sec:5.3} we discuss possible origins of the absorbing
gas.

\subsection{The EW host galaxy mass relation}

We here assume that all the \mgii\ absorption lines detected are associated 
to the gaseous halo of the QSOs. However, we can not exclude the possibility 
that some of these systems are due to intervening galaxies (see~\ref{sec:5.3}).

High resolution observations show that the \mgii\ absorbers 
associated to galaxies are often splitted in several discrete 
components \citep[e.g.,][]{Churchill2001}, suggesting a 
clumpy nature of the gaseous halo. Thus the \ewr\ of an 
absorption line is roughly proportional to the number of absorbing 
component along the line of sight \citep[e.g.,][]{Petitjean1990, 
Churchill2003}.
 
\begin{figure}
\centering
\includegraphics[width=1.0\columnwidth]{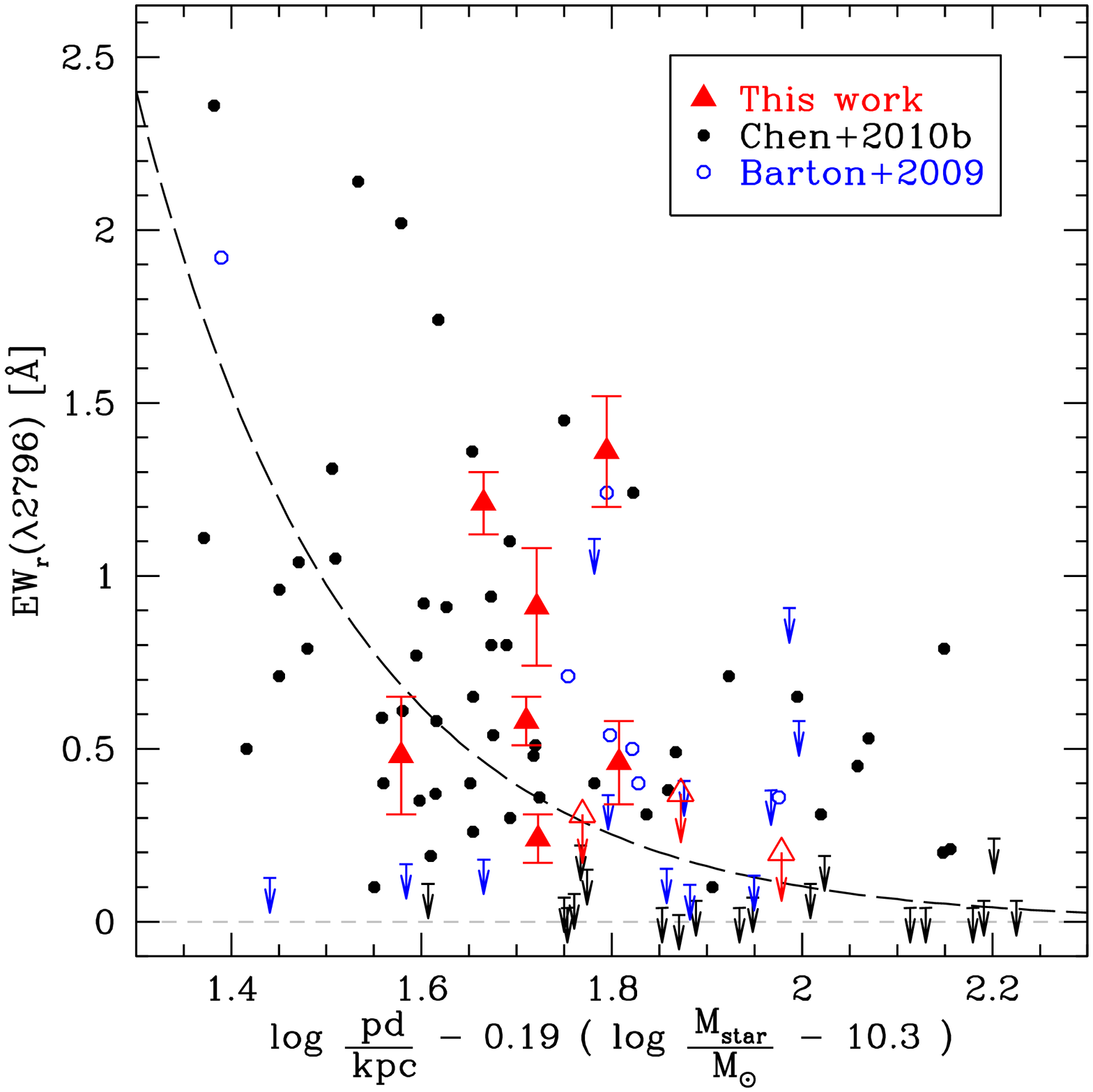}
\caption{Rest frame EW of \mgii($\lambda2796$) absorption line as a function 
of projected distance and stellar mass for QSOs and galaxies. Red filled 
triangles are systems for which the absorption system was detected, while the 
empty ones are 2$\sigma$ upper limits. Empty blue and filled black 
circle are data for galaxies from \citet[][with upper limits converted 
to the considered 2$\sigma$ values]{Barton2009} and \citet{Chen2010b}, respectively. Black 
dashed line is the \ewr\ vs. 
projected distance anti--correlation for galaxies including the scaling
relation with stellar mass proposed by \citet{Chen2010b}.
For the x--axis we have adopted the same projection of Figure~3 in 
\citet{Chen2010b}.}\label{fig:6}
\end{figure}

For the sake of simplicity we here assume that the \ewr\ reflects 
the potential well of the dark matter halo and thus that the 
distribution of the cool gas around galaxies follow a spherical
profile \citep[e.g.,][]{Srianand1993, Srianand1994, Tinker2008, Chelouche2008}. 
At a given radius, the more massive systems sustain larger gaseous haloes 
that are responsible for stronger absorption lines due to the larger number 
of clouds intercepted. 
Assuming that the mass of the QSO host galaxy ($\mhost$) trace the mass of 
the extend dark matter halo \citep[see e.g.,][]{More2011}, despite the 
uncertainties on the mass--to--light ratio, the \ewr\ of the absorption features is 
expected to follow a relation:
\begin{equation}
{\rm EW}_{\rm r} \propto \frac{{\rm M}_{\rm host}^{\alpha}}{{\rm pd}^{\beta}}.
\end{equation}
For instance, the two coefficients $\alpha$ and $\beta$ have been estimated by 
\citet{Chen2010b} in a sample of low redshift galaxies ($\z < 0.5$):
$\alpha=1.8\pm0.1$ and $\beta=0.34\pm0.06$ (see Figure~\ref{fig:6}).  

Is is worth of noting that the correlation between the \ewr\ and the mass 
of the halo is still not well established. For instance, \citet{Bouche2006}, 
from an analysis of the relation between \mgii\ absorbers and luminous red 
galaxies in the SDSS found an anti--correlation between the absorber halo mass 
and \ewr, i.e., on average very strong absorbers (\ewr$\gsim2$\,\AA) arise 
in less massive dark matter haloes ($\sim10^{11}\msun$) while absorbers with 
\ewr$\sim0.3-1.2$\,\AA\ in the more massive one ($\sim10^{12.5}\msun$). 
In addition \citet{Charlton1996} show that the \mgii\ distribution could be  
satisfactorily explained with both an extended disk geometry as well as a 
spherical one. From a sample of 40 galaxies at redshift $\sim0.5$,
\citet{Kacprzak2011} suggest that the \mgii\ gas is distributed following a 
flattened halo that is co--planer and coupled to the inclination of the 
galaxy disc \citep[see also,][]{Kacprzak2012}.

In order to determine the mass of the host galaxy ($\mhost$) of \qsof\ we 
consider the $\mbh$--$\mhost$ relation presented by \citet{Decarli2010b}
that is based on the investigation of 96 QSOs with known host galaxy 
luminosities in the redshift range $0.07<\z<2.74$:
\begin{equation}
\log{\frac{\mbh}{\mhost}} = \left(0.28\pm0.06\right)\z - \left(2.91\pm0.06\right)
\end{equation}
where $\z$ is the QSO redshift. 
The $\mbh$s are estimated using the virial method applied to the gas of the 
BLR. These could be inferred from the width of broad emission lines and from 
the continuum luminosity, as expected from a photoionisation model 
\citep[e.g.,][]{Kaspi2000}. The uncertainties associated to the $\mbh$
estimate are dominated by the dispersion of the relation between the radius of 
the BLR and the luminosity of the continuum and are typically around $\sim0.4$\,dex
\citep[e.g., ][]{Vestergaard2006, Shen2011b}. The $\mbh$ calculated from the recipes 
of~\citet{Vestergaard2006} and of~\citet{Vestergaard2009} are in Table~\ref{tab:1}.

In Figure~\ref{fig:6} we show the distribution of the \ewr\ as a 
function of the impact parameter rescaled for the stellar mass for
quiescent galaxies from \citet{Barton2009} and \citet{Chen2010b} 
(on average $\mgal\sim10^{10}\msun$) and for the host of QSOs
(on average $\mhost\sim2\times10^{11}\msun$). 
Taking into account the mass, the \mgii\ absorption systems associated 
to QSOs have a distribution close to that of galaxies. The value of the 
$\chi^2$ for our data estimated on the relations presented by 
\citet{Chen2010a, Chen2010b} prior and after accounting for the galaxy 
mass improves of a factor $\sim3$. 
This is in agreement with studies that suggest that the haloes of QSOs are 
similar to that of normal galaxies \citep[e.g.,][]{Chelouche2008}.

\subsection{Effects of the QSO radiation}

The presence of the intense radiation field coming from
the central supermassive black hole should have a substantial effect in 
the thermal state of the gaseous halo. \citet{Chelouche2008} model the 
distribution of cool gas around QSOs suggesting that it is filled
with clouds having size of $\sim1$\,pc and density of 
$\sim10^{-2}$\,cm$^{-3}$. Under these conditions a QSOs with luminosity
of $10^{46}$\,erg/s \citep[the average of our sample calculated from 
the bolometric corrections given in][see Table~\ref{tab:1}]{Richards2006} 
can heat the cool gas  within $\sim100$\,kpc up to a temperature of 
T$\sim10^5$\,K, allowing the existence of only few \mgii\ absorbers. 
\citet{Wild2008} from a sample \mgii\ absorber with equivalent
width larger than $0.3$\,\AA\ directly associated to the QSOs, found
that the QSO emission destroys \mgii\ clouds out to at least
$800$\,kpc.
Since the QSO radiation is thought to be emitted into cones 
\citep[][]{Antonucci1993}, an anisotropy of the distribution 
of absorbers is thus expected \citep[e.g.,][]{Hennawi2006}.
If the UV emission of the central black occurs along the line of sight,
the transverse absorption features will be not affected by it, and the QSO 
luminosity will have little impact on the extend \mgii\ absorbing gas 
at large galactic radii. On the contrary, \mgii\ absorbers along the line
of sight are photoionised by the QSO emission \citep[see also,][]{Elvis2000}.

\civ\ and \mgii\ are believed to reside in the same halo clouds, 
however the ionising region of the former is expected to be smaller than
that of the latter, consistent with its higher ionisation energy. 
\civ\ absorbers could survive within $\sim100$\,kpc,
even though the radiation field of the QSO \citep{Chelouche2008}.
The origin of the \civ\ LOS absorbers could also be associated to
the region closer to the QSO. For instance, \citet{Wild2008}, investigating 
the \civ\ LOS absorption system in a sample of $\sim7400$ \civ\ absorbers from 
the SDSS, have found that $\sim40\%$ of the systems within $3000$\,km/s of 
the QSO are due to a gas outflow from the central black hole 
\citep[see also,][]{Nestor2005}. 

\subsection{Origin of Mg\,II absorbing gas}\label{sec:5.3}

Although our sample of 13 QSO pairs does not allow us to put firm 
constraints, we can comment on the possible origin of the absorbing gas.

A simple explanation for the most intense and distant absorption lines 
detected around QSOs is that these are due to a chance superposition of 
a satellite galaxy or to the presence of tidal tail due to recent merger 
episodes \citep[see e.g.,][]{Fumagalli2011, Keeney2011}. This scenario 
is supported by studies suggesting that, despite the wide variety of 
environments in which they are detected, QSOs seems to prefer galactic 
environment richer than average \citep[e.g.,][]{Wold2001, Serber2006, 
Hutchings2009}. However, the high  covering factor and the observational 
evidence of galaxies surrounded by cool gas up to a radius of $\sim100$\,kpc, 
allow us to fairly assume the link between absorbers and QSO haloes. In 
support of this we also roughly estimate the the \hi\ column density from 
the rest frame equivalent width of \mgii\ from the relations found
in \citet{Menard2009}. All the detected 
systems have $\log \left(N_{\rm HI}/{\rm cm}^2\right)< 19.6$, 
that is unlikely associated to galactic disc \citep[e.g.,][]{Zwaan2005}. 
However, we can not exclude the possibility that the absorptions origin 
from extra--planar neutral gas associated to spiral galaxies 
\citep[e.g.,][]{Sancisi2008}.

If the absorbers are associated to the QSO gaseous halos, the improvement 
of the anti--correlation between \ewr\ and projected distance due to the addition 
of the galaxy mass as parameter is naturally explained in the inflow scenario. 
As suggested by \citet{Chen2010b} more massive galaxies have more extended 
halos of cool gas and thus the observed \mgii\ absorbers arise in infalling 
clouds. However, we note that only the weaker absorption systems are thought 
to be associated to inflows of gas (the few absorption systems observed by 
Chen et al. at separations larger than 70\,kpc have typically 
\ewr$\lsim0.7$\,\AA), while the stronger ones to outflows 
\citep[e.g.,][]{Zibetti2007}. 
The  inflows of gas responsible of the QSO activity may produce intense
star formation episodes and strong winds \citep[e.g.,][]{Dimatteo2005, 
Hopkins2005, Maiolino2012} that might give rise to the stronger detected 
absorption lines. 
This is also supported by the recent discovery of signature of high star 
formation rate and gas outflows in the QSO host galaxies \citep[][]{Floyd2012, 
Sanmartim2012}.

\section[]{Summary and Conclusions}\label{sec:6}

We studied 13 close projected QSO pairs observed with FORS2 at ESO--VLT. 
The projected separations of the systems (between $60$\,kpc and $120$\,kpc) 
allowed us to to investigate the presence of \mgii\ and \civ\ absorption
systems in the outer regions of the extended gaseous haloes of QSOs in the 
redshift range $0.7\lsim \zf \lsim 2.2$.

In 7 out of 10 systems we have detected the \mgii\ doublet on the \qsob\ 
spectrum in correspondence to the \qsof, while only one association out of 3 
is found for \civ.
We compare our results with those performed for normal galaxies by 
\citet{Kacprzak2011}, \citet{Chen2010a}, and \citet{Barton2009}.
If we consider \ewr\ as a function of both the projected distance and the 
mass of the systems, we find that the haloes of QSOs are similar to that of 
galaxies. In our small sample, we do not detect a significant
enhancement of the absorption system strengths, as could be expected if the QSO 
nuclear activity were driven by intense gas accretion onto the black hole.
Finally, we note that along the line of sight we do not detect any \mgii\ 
absorbers of the same strength of the transverse one.
These results are in agreement with models that consider a non 
isotropic emission of the QSO, which are hosted by gaseous
haloes more massive that those of galaxies.

\section*{Acknowledgements}

We acknowledge helpful discussions with C.~Montuori, F.~Haardt, and 
M.~Fumagalli.
For this work EPF was supported by Societ{\`a} Carlo Gavazzi S.p.A. and by 
Thales Alenia Space Italia SpA. RD acknowledges funding from Germany's
national research centre for aeronautics and space (DLR, project FKZ 50
OR 1104). This work was based on observations made with the ESO/VLT 
Telescope in Paranal.

\appendix

\section[]{Notes on individual objects}\label{app:1}

\noindent {\bf QQ01} --- 
The \mgii\ absorption lines in QQ01B is detected at a level slighty lower 
than the considered 3$\sigma$ threshold. However, since the large uncertainties 
in the continuum location, we have maintained this absorption system in the 
analysis.

\noindent {\bf QQ04} --- 
The two QSOs of the pair are very close in redshift thus the association of 
the absorption feature to the foreground or to the background QSO is uncertain.
QQ04F shows two \civ\ absorption systems superimposed to the 
broad emission line. 
A stronger one at redshift $\z_{\rm abs}\sim 2.1120$ and a fainter 
blueshifted one at $\z_{\rm abs}\sim 2.1350$. The two component of 
the stronger absorption system are partially blended, and thus fitted 
simultaneously with the sum of two Gauss functions. Moreover, other 
absorptions lines are present at $\lambda=4822.7$\,\AA\ and at 
$\lambda=4845.8$\,\AA, whose identification is uncertain as they are 
superimposed on the stronger absorption system.

Due to the velocity range that we have considered to associate the absorption 
features to the \qsof\ (see~\S\ref{sec:4}), we have assumed that the 
stronger feature is due to the QSO halo, while the weaker one (with $\Delta V\sim3000$\,km/s
from the former) to a simple chance superposition. However, the uncertainties
in the redshift determination based on \civ\ broad emission line are large,
mostly due to the presence of the strong absorption features 
\citep[see e.g.,][]{Hewett2010}. 
This does not allow to exclude the possibility that the weaker absorption 
system is connected to the \qsof\ halo and that the stronger one is due to 
a gas outflow.
QQ04B shows a broad absorption lines in the blue wing of the \civ\ emission 
lines. These P--Cygni like profile, are usually interpreted 
as radiatively driven winds associated to strong outflows from QSOs 
\citep[e.g.,][]{Scargle1970, Turnshek1984}. In the spectra of QQ04B, two more 
\civ\ absorption features are present at redshift z$_{\rm abs} = 2.0302$ and 
z$_{\rm abs} = 2.0443$, suggesting a very rich environment for this pair 
or an interaction between the QSOs (see Table~\ref{tab:3}).

\noindent {\bf QSO05} --- 
In the spectra of QQ05B we detect \feii($\lambda$2586) and \feii($\lambda$2600) lines at the
same redshift of the \mgii\ doublet associated to the \qsof\
(\ewr($\lambda$2586)$=(0.12\pm0.04)$\,\AA\ and \ewr($\lambda$2600)$=(0.39\pm0.04)$\,\AA).

\noindent {\bf QSO07} --- 
Associated to the \mgii\ absorption feature at redshift $\z_{\rm abs}=1.3355$
(see Table~\ref{tab:3}) we detect \feii($\lambda$2586) and \feii($\lambda$2600) lines with
\ewr($\lambda$2586)$=(0.61\pm0.07)$\,\AA\ and \ewr($\lambda$2600)$=(0.70\pm0.04)$\,\AA. 
A \feii($\lambda$2382) line is present at the same redshift of the \mgii\ 
doublet at $\z_{\rm abs}=1.5186$
(\ewr($\lambda$2382)$=(0.23\pm0.04)$\,\AA).

\noindent {\bf QSO10} --- 
Two \civ\ doublets are superimposed to the \civ\ broad emission of \qsob.
Their redshifts are $\z_{\rm abs}= 2.5127$ and $\z_{\rm abs}= 2.5167$, 
corresponding to a velocity difference of $\sim500$\,km/s (see 
Table~\ref{tab:3}). Due to the large uncertainties in the QSO 
redshift determination from broad emission lines (see \S\ref{sec:4}), 
both the systems could be associated to the QSO.

\noindent {\bf QSO13} ---
This pair was already investigated by \citet{Decarli2009}, which have noticed 
a non resolved \mgii\ absorption feature associated to the foreground QSO. 
Moreover they observe QSO13F in Ks band to detect the host galaxy. This allows 
to give a indipendent estimate of the host galaxy mass 
(${M}_{\rm host}=5\times10^{11}\,\msun$) that is consistent with the 
value estimated indirectly from the $\mbh$. 

In the spectra of \qsob\ we identify two \civ\ doublet at redshifts 
discordant with that of the QSO ($\z_{\rm abs}\sim2.3215 $ and 
$\z_{\rm abs}\sim 2.3272$). Due to their redshift difference 
the component of the doublets are superimposed, we thus infer the \ewr\
assuming that the lines are not saturated and thus a theoretical doublet 
ratio value of DR$=2$ 
(see Table~\ref{tab:3}).

\begin{table*}
\centering
\caption{Properties of \mgii\ and \civ\ absorption features not associated to \qsof\ 
detected on the QSO spectra:
our identification label of the QSO (ID), 
feature detected (Abs.),
observed wavelength ($\lambda_{\rm abs}$),
rest frame equivalent width (EW$_{\rm r}$),
doublet ratio (DR), and
redshift (z$_{\rm abs}$).
}\label{tab:3}
\begin{tabular}{lcccccc}
\bf{Mg\,II} &&&&& \\      	  
\hline
ID     &  $\lambda_{\rm abs}(\lambda2796)$ & EW$_{\rm r}(\lambda2796)$ &  $\lambda_{\rm abs}(\lambda2803)$ & EW$_{\rm r}(\lambda2803)$ & DR & z$_{\rm abs}$ \\ 
       &  [\AA]	                           & [\AA]        	       &  [\AA] 			   & [\AA]                     &    &              \\      	  
\hline  		 
\hline  		 
 QQ01F & 5954.7  & 0.86$\pm$0.15 & 5970.3  & 0.45$\pm$0.14 & 1.91$\pm$0.68\phantom{$^{\text \textdagger}$} & 1.1294$\pm$0.0001 \\ 
 QQ07B & 6378.1  & 0.46$\pm$0.06 & 6395.0  & 0.43$\pm$0.03 & 1.07$\pm$0.16\phantom{$^{\text \textdagger}$} & 1.2810$\pm$0.0002 \\
 QQ07B & 6531.0  & 1.10$\pm$0.04 & 6547.9  & 1.09$\pm$0.06 & 1.01$\pm$0.07\phantom{$^{\text \textdagger}$} & 1.3355$\pm$0.0001 \\
 QQ07B & 7042.9  & 0.58$\pm$0.06 & 7061.0  & 0.46$\pm$0.04 & 1.26$\pm$0.17\phantom{$^{\text \textdagger}$} & 1.5186$\pm$0.0002 \\  
 QQ08B & 6958.9  & 1.02$\pm$0.10 & 6975.6  & 0.51$\pm$0.11 & 2.00$\pm$0.47\phantom{$^{\text \textdagger}$} & 1.4882$\pm$0.0008 \\
\hline 
\hline
         &&&&& \\      	  
\bf{C\,IV} &&&&& \\      	  
\hline
ID     &  $\lambda_{\rm abs}(\lambda1548)$ & EW$_{\rm r}(\lambda1548)$ &  $\lambda_{\rm abs}(\lambda1551)$ & EW$_{\rm r}(\lambda1551)$ & DR & z$_{\rm abs}$ \\ 
       &  [\AA]	                           & [\AA]        	       &  [\AA] 			   & [\AA]                     &    &   	    \\         
\hline  		 
\hline 
 QQ04B & 4691.5  & 0.54$\pm$0.03 & 4699.1  & 0.37$\pm$0.02 & 1.46$\pm$0.11\phantom{$^{\text \textdagger}$} & 2.0302$\pm$0.0002 \\
 QQ04B & 4713.4  & 0.27$\pm$0.02 & 4721.0  & 0.22$\pm$0.05 & 1.23$\pm$0.29\phantom{$^{\text \textdagger}$} & 2.0443$\pm$0.0003 \\
 QQ10B & 5438.4  & 0.22$\pm$0.03 & 5447.4  & 0.16$\pm$0.06 & 1.38$\pm$0.55\phantom{$^{\text \textdagger}$} & 2.5127$\pm$0.0002 \\
 QQ10B & 5444.7  & 0.19$\pm$0.06 & 5453.6  & 0.19$\pm$0.04 & 1.00$\pm$0.38\phantom{$^{\text \textdagger}$} & 2.5167$\pm$0.0002 \\
 QQ13B & 5142.3  & 0.14$\pm$0.03 & 5151.1  & 0.07$\pm$0.03 & 2.00$\pm$0.96$^{\text \textdagger}$           & 2.3215$\pm$0.0001 \\
 QQ13B & 5151.1  & 0.40$\pm$0.03 & 5159.7  & 0.20$\pm$0.03 & 2.00$\pm$0.34$^{\text \textdagger}$           & 2.3272$\pm$0.0001 \\
\hline 
\hline
\multicolumn{7}{l}{$^{\text \textdagger}$ Assumed value for DR (see text for details)} \\      	    										       
\end{tabular}
\end{table*}

\label{lastpage}
\end{document}